\documentclass[copyright,creativecommons]{eptcs}
\usepackage{graphicx, preamble}

\usepackage{quiver}

\newcommand{\vicartitle}{\vicar: Visualizing Categories with Automated Rewriting in Coq}

\title{\vicartitle}
\author{
Bhakti Shah*\blfootnote{Equal contribution} \quad William Spencer* \quad Laura Zielinski* \\ Ben Caldwell \quad Adrian Lehmann \quad Robert Rand
\institute{University of Chicago\\
Chicago, USA}
}
\def\titlerunning{\vicartitle}
\def\authorrunning{Shah, Spencer, Zielinski, Caldwell, Lehmann \& Rand}
\providecommand{\event}{ACT 2024}

\begin{document}
\bibliographystyle{eptcs}
\maketitle
\begin{abstract}
\textbf{Abstract.} We present \vicar, a library for working with monoidal categories in the Coq proof assistant. \vicar provides definitions for categorical structures that users can instantiate with their own verification projects. Upon verifying relevant coherence conditions, \vicar gives a set of lemmas and tactics for manipulating categorical structures. We also provide a visualizer that can display any composition and tensor product of morphisms as a string diagram, showing its categorical structure. This enables graphical reasoning and automated rewriting for Coq projects with monoidal structures. %
\end{abstract}

\section{Introduction}\label{sec:Introduction}
Just as category theory provides a unifying framework for diverse concepts in mathematics, category theory can be used in formal verification as a basis for generalization. 
We can abstract away common structural patterns to yield reusable tactics, lemmas, and techniques for proof assistants. 
Many constructs that appear in program verification resemble symmetric monoidal categories.
\vicar, a library for Visualizing Categories with Automated Rewriting, takes advantage of the shared structure across symmetric monoidal instances to understand and manipulate morphisms in Coq. 

\vicar emerged from \vyzx, an effort to verify the \zxcalc, a graphical reasoning system for quantum programs~\cite{2023vyzx}. The \zxcalc~\cite{coecke-duncan-zx} forms a symmetric monoidal category whose morphisms are the \zxdiags which make up the language. \vicar generalizes the tools developed in \vyzx to assist Coq verification for all concrete monoidal categories. \vicar consists of three parts: typeclasses for categorical structures in Coq, a visualizer to represent morphisms using string diagrams, and a set of tactics for manipulating typeclass instances.

Other popular examples of monoidal categories include the calculus of relations and matrices. The former is the category whose objects are types and morphisms are binary relations. 
The latter has matrices as its morphisms between vector spaces.
The similarities between these examples are initially unclear. Reframing each categorically, however, reveals their shared structure and hints at how the verification of one could help another. To explore and justify this claim, we imported or implemented these instances in Coq then applied \vicar's monoidal category framework. We found that for a well-developed project, we could easily instantiate the relevant categorical definitions. The tactics and visualization gained were valuable and helped project-specific proofs by removing proof and cognitive overhead. We discuss these examples in more detail in \Cref{sec:examples}.

\vicar's key contributions are as follows:
\begin{itemize}
    \item We define symmetric monoidal categories in Coq, easily instantiable by user-created structures.
    \item We present an automatic morphism visualizer. While working on a Coq proof, the visualizer will parse the monoidal structure of the current proof state, producing an image of its string diagram representation.
    \item We provide a set of powerful automation tactics that users can access once they have proven the necessary coherence conditions. They include \coqe{foliate}, which automatically rewrites diagrams to common useful structures; \coqe{assoc_rw}, which performs rewriting modulo associativity; and \coqe{cat_simpl}, a simplification tactic for commonly occurring patterns.
    \item We formalize the coherence theorem for monoidal categories and develop from it a powerful tactic, \coqe{monoidal}, which employs proof by reflection.
\end{itemize}

We describe \vicar's contributions in detail, how to use them, and how they fit into \vyzx and other projects with categorical structure. We conclude the paper with some discussion of the next steps for \vicar. Our work is open source and available on GitHub: \url{https://github.com/inQWIRE/ViCAR}.

\section{Background}\label{sec:Background}

\paragraph{Formal verification and the Coq proof assistant}\label{subsec:FVandCoq}
Formal verification is grounded in the idea that we can mathematically prove that a computer program satisfies a specification. 
This guarantees that a given piece of software performs as expected on all possible inputs. Formal verification research focuses on developing new and more effective techniques for these kinds of proofs. 
One approach uses proof assistants, which are software (usually programming languages) that allow users to express and prove various constructs, often mathematical, in code. 

Among the most widely-used proof assistants is Coq~\cite{Coq12}.
It allows for writing definitions in the style of a dependently-typed programming language and proving statements in the style of mathematical proof.
One powerful Coq mechanism is custom tactics, written in the Ltac language, which automate repetitive tasks across proofs. 
Often, Coq libraries offer a set of tactics to abstract away technical details and perform complex actions. 
\vicar aims to provide such tactics for users verifying projects which have categorical structure.

\paragraph{Symmetric monoidal categories}\label{subsec:SymmetricMonoidal}

\begin{figure}[b]
    \centering
    \begin{tikzcd}[column sep=tiny]
    	{(A\otimes 1)\otimes B} && {A\otimes(1\otimes B)} && {(A\otimes B)\otimes (C\otimes D)} \\
    	&&& {((A\otimes B)\otimes C)\otimes D} && {A\otimes (B \otimes (C \otimes D))} \\
    	& {A\otimes B} && {(A\otimes (B\otimes C))\otimes D} && {A\otimes ((B\otimes C)\otimes D)}
    	\arrow["{\alpha_{A,1,B}}", from=1-1, to=1-3]
    	\arrow["{\mathrm{id}_A\otimes\lambda_B}", from=1-3, to=3-2]
    	\arrow["{\rho_A\otimes\mathrm{id}_B}"', from=1-1, to=3-2]
    	\arrow["{\alpha_{A\otimes B,C,D}}"{pos=0.3}, from=2-4, to=1-5]
    	\arrow["{\alpha_{A, B,C\otimes D}}"{pos=0.7}, from=1-5, to=2-6]
    	\arrow["{\alpha_{A,B,C}\otimes \mathrm{id}_D}"', from=2-4, to=3-4]
    	\arrow["{\alpha_{A,B\otimes C,D}}"', from=3-4, to=3-6]
    	\arrow["{\mathrm{id}_A \otimes \alpha_{B,C,D}}"', from=3-6, to=2-6]
    \end{tikzcd}
\caption{The triangle identity (left) and the pentagon identity (right)}\label{fig:diagrams}
\end{figure}

Following the definition of Selinger~\cite{Selinger2010}, a (planar) monoidal category consists of a base category $\catC$ equipped with a bifunctor $\otimes:\catC \times \catC \rightarrow \catC$ called the tensor product. The tensor product is required to be unital and associative, which means there are natural isomorphisms whose components are, for objects $A,B,C\in\catC$,
\begin{align}
    \lambda_A&:1\otimes A \xrightarrow{\sim} A, \tag{left unitor} \\
    \rho_A&:A \otimes 1 \xrightarrow{\sim} A, \tag{right unitor} \\
    \alpha_{A,B,C}&:(A \otimes B) \otimes C \xrightarrow{\sim} A \otimes (B \otimes C). \tag{associator}
\end{align}

These natural isomorphisms are required to be coherent, meaning any diagram of a certain formal class made only of associators and unitors must commute (for details, see Selinger~\cite{Selinger2010}). 
It has been proven that showing commutativity of just two types of diagrams is sufficient to ensure this powerful result~\cite{eilenberg1966closed}.  Specifically, the natural isomorphisms of a monoidal category $\catC$ are coherent if and only if the diagrams in \Cref{fig:diagrams} commute for all $A, B, C, D \in \catC$. This result is called the coherence theorem for monoidal categories, which we may also refer to as monoidal coherence~\cite{eilenberg1966closed}.
\vicar includes a formalization of this theorem and uses it to create the \coqe{monoidal} tactic, as we explain in \Cref{sec:automation}. 

A monoidal category can further be braided, meaning that there is a natural isomorphism whose components are, for objects $A,B\in\catC$, 
\begin{equation*}
    \tag{braiding}
    \beta_{A,B}:A\otimes B \xrightarrow{\sim} B\otimes A.
\end{equation*}
Again, we require that two diagrams called the hexagon identities commute, which ensure the coherence of the braiding.
Finally, a braided monoidal category is symmetric if, for all $A,B\in\catC$,
\begin{equation}
    \beta_{A,B}\circ \beta_{B,A}\simeq 1.\tag{symmetry}
\end{equation}
Essentially, symmetric monoidal categories have an almost-commutative tensor product. We choose to focus on monoidal and symmetric monoidal categories because of their prevalence and natural correspondence to string diagrams \cite{Selinger2010}. String diagrams are a graphical way to represent morphisms, useful for parsing complex structure. One of our goals is to unify verification, which is almost always text-based, with visual reasoning. String diagrams are the platform for doing so for monoidal categories. We define them concretely in \Cref{sec:viz}.

\paragraph{Categories in verification} \label{subsec:CategoriesInVerification}
In mathematics and computer science, symmetric monoidal categories are everywhere.
Commonly seen examples include the category of sets, the category of finite vector spaces, and the simply-typed lambda calculus. In verification, active projects whose core structure is a symmetric monoidal category include the verification of the \zxcalc~\cite{2023vyzx} and that of causal separation diagrams~\cite{castello2023inductive}. 
The \zxcalc is a graphical language for expressing quantum computation, while causal separation diagrams allow us to reason about parallel processes. Both constructs independently satisfy the definitions of a symmetric monoidal category, though this fact is not directly used in their verification.

Despite being in completely different domains, because of their shared structure, the \zxcalc and causal separation diagram have properties in common that should be exploited to ease their verification. This is the idea that inspired \vicar and the gap in formal verification that we wanted to address\footnote{We note that VyZX and causal separation diagrams are implemented in different proof assistants, Coq and Agda, respectively. We chose to use Coq for \vicar.}. We bridge this gap with our framework for instantiating monoidal categories, our generalized rewriting tactics, and our morphism visualizer. We prioritize automation and ease---we want users to be able to use \vicar's features in their own proofs with minimal additional effort. 

Other projects attempt to unify categorical reasoning, proof assistants, and visualization. 
The \chyp proof assistant, for one, allows users to state rewrite rules axiomatically and produces string diagrams to visualize the rules in action~\cite{chyp}. 
\vicar takes the alternative approach of requiring users to prove their structures are instances of predefined category typeclasses. 
In exchange for this effort, \vicar proofs can be used within the greater context of the Coq proof assistant and augment existing Coq projects. 
\vicar's approach allows us to use categorical reasoning to verify existing software, while \chyp is able to more easily handle rewrites modulo associativity.

\section{Constructively defining categories in Coq}

There are several preexisting examples of implementing category theory in Coq~\cite{gross2014experience,jwiegleycategory}. 
Though many of these libraries have significant developments, we found they did not align with all of our goals.
For instance, we want to separate structural definitions from coherence conditions to enable easy visualization of category instances, independent of their semantics.
Moreover, when working with categories in practice, some notion of morphism equivalence is assumed, so we benefit by making this explicit in formal verification.
For \vicar, we are interested in category theory as a means to generalize shared structure. Our library is to be instantiated by active verification projects across a range of domains. We made a number of technical decisions to reflect this purpose.

We implement our categorical definitions using a hierarchy of Coq typeclasses~\cite{coq-typeclasses}, a mechanism similar to interfaces in object-oriented programming. Typeclasses specify and label a collection of types, possibly dependent on each other, and instances of that typeclass must provide a concrete term for each type. For example, \Cref{fig:coherence} gives the part of the typeclass for monoidal categories which translates the coherence conditions from \Cref{fig:diagrams}. Typeclasses can also inherit from each other. \vicar's typeclass hierarchy starts with the base category, then monoidal category, braided monoidal, and finally symmetric monoidal. A project may instantiate as many typeclasses as is suitable for its purposes. 

\begin{lstlisting}[caption=The triangle and pentagon identities in the \coqe{MonoidalCategoryCoherence} typeclass.,label=fig:coherence,
float=tp,floatplacement=tbp %
]
Class MonoidalCategoryCoherence {C : Type} {cC : Category C}
  {cCh : CategoryCoherence cC} (mC : MonoidalCategory cC) : Type := {
    triangle (A B : C) : 
        α_ A, I, B ∘ (id_ A ⊗ λ_ B)
        ≃ ρ_ A ⊗ id_ B;
    pentagon (A B M N : C) : 
        (α_ A, B, M ⊗ id_ N) ∘ α_ A, (B × M), N ∘ (id_ A ⊗ α_ B, M, N)
        ≃ α_ (A × B), M, N ∘ α_ A, B, (M × N);
    (* Remainder omitted *)
}.
\end{lstlisting}

A benefit of using typeclasses is Coq's inference mechanism, which automatically searches for typeclass instances~\cite{coq-typeclasses}. This avoids having to reference a particular instance every time one of its terms is used. For example, suppose an instance \coqe{catC} of our category typeclass has been declared whose objects have type $C$. Then, if $A, B$ and $M$ are terms of type $C$ and \coqe{f} and \coqe{g} have types $A\leadsto B$ and $B\leadsto M$, the expression \coqe{f $\circ$ g} would type check without the user having to explicitly point to the particular instance. Coq would automatically retrieve it, determining that morphism composition $\circ$ should be taken as defined by \coqe{catC}.

One important way \vicar differs from existing formalizations of category theory is its separation of structural definitions from coherence conditions. Each abstract category has a typeclass containing just the necessary structures, such as identity or the associator, and a different typeclass specifying its coherence conditions. There are two separate hierarchies accordingly. The benefit is that users can access the visualizer without having to prove coherence, which may be more demanding. Instantiating a structural typeclass with existing definitions is enough to begin using our visualizer. Of course, without proving coherence, one has no guarantee that the provided structure actually satisfies the criteria of an abstract category, and therefore cannot use our automation.

Another way \vicar diverges from other formalizations is by requiring users to supply an explicit equivalence relation for morphisms, denoted $\simeq$, instead of always using built-in equality.
Using such an equivalence relation is necessary for many standard constructions because Coq does not support performing a quotient by an equivalence relation. 
By working over the supplied morphism equivalence, we maintain this flexibility that many implementations rely on. 
We also chose to use a diagrammatic compose for the notation $\circ$, as our project focuses on visualization. 

\section{Visualization}
\label{sec:viz}

\paragraph{Categorical String Diagrams}
Reasoning about morphisms in categories is assisted by the use of string diagrams, an associated graphical language. String diagrams visually represent monoidal categories that are ordinarily represented in text form.
Their notation focuses on morphisms rather than objects, and can therefore very concisely represent complicated expressions.
We visualize our categories as string diagrams to reduce cognitive overhead during proof.

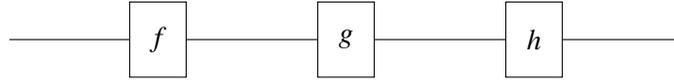
\begin{figure}
    \centering
    \begin{tikzpicture}
        \begin{pgfonlayer}{nodelayer}
            \node [style=medium box] (0) at (-2.5, 0) {$f$};
            \node [style=medium box] (1) at (0, 0) {$g$};
            \node [style=medium box] (2) at (2.5, 0) {$h$};
            \node [style=none] (3) at (-4.5, 0) {};
            \node [style=none] (4) at (4.5, 0) {};
        \end{pgfonlayer}
        \begin{pgfonlayer}{edgelayer}
            \draw (3) to (0);
            \draw (0) to (1);
            \draw (1) to (2);
            \draw (2) to (4);
        \end{pgfonlayer}
    \end{tikzpicture}
    \caption{Visualization of both $f\circ (g\circ h)$ and $(f\circ g)\circ h$ in traditional string diagrams.}
    \label{fig:assoc-example-a}
\end{figure}

\begin{figure}
\centering
\begin{subfigure}{.45\textwidth}
  \centering
  \includegraphics[width=0.9\linewidth]{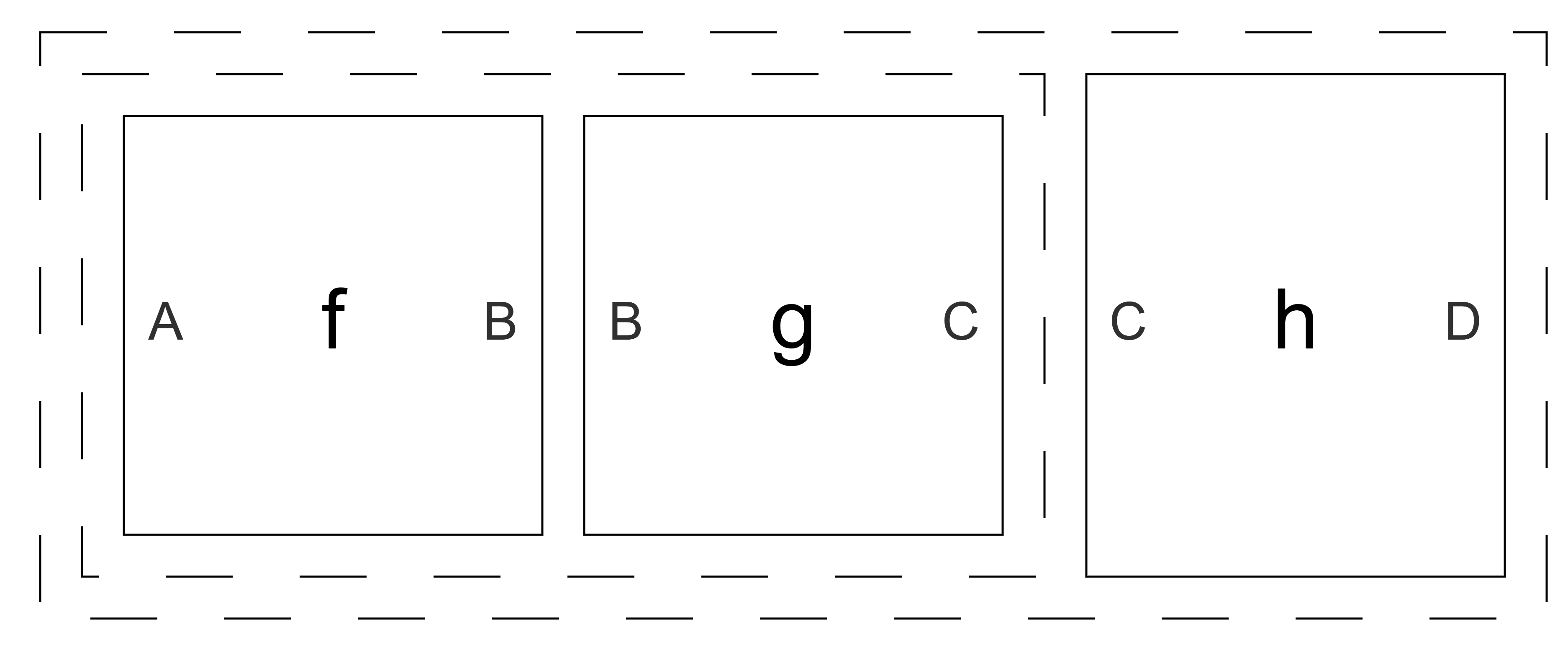}
  \caption{\vicar visualization of \coqe{(f ∘ g) ∘ h}.}
\end{subfigure}
\begin{subfigure}{.45\textwidth}
  \centering
  \includegraphics[width=0.9\linewidth]{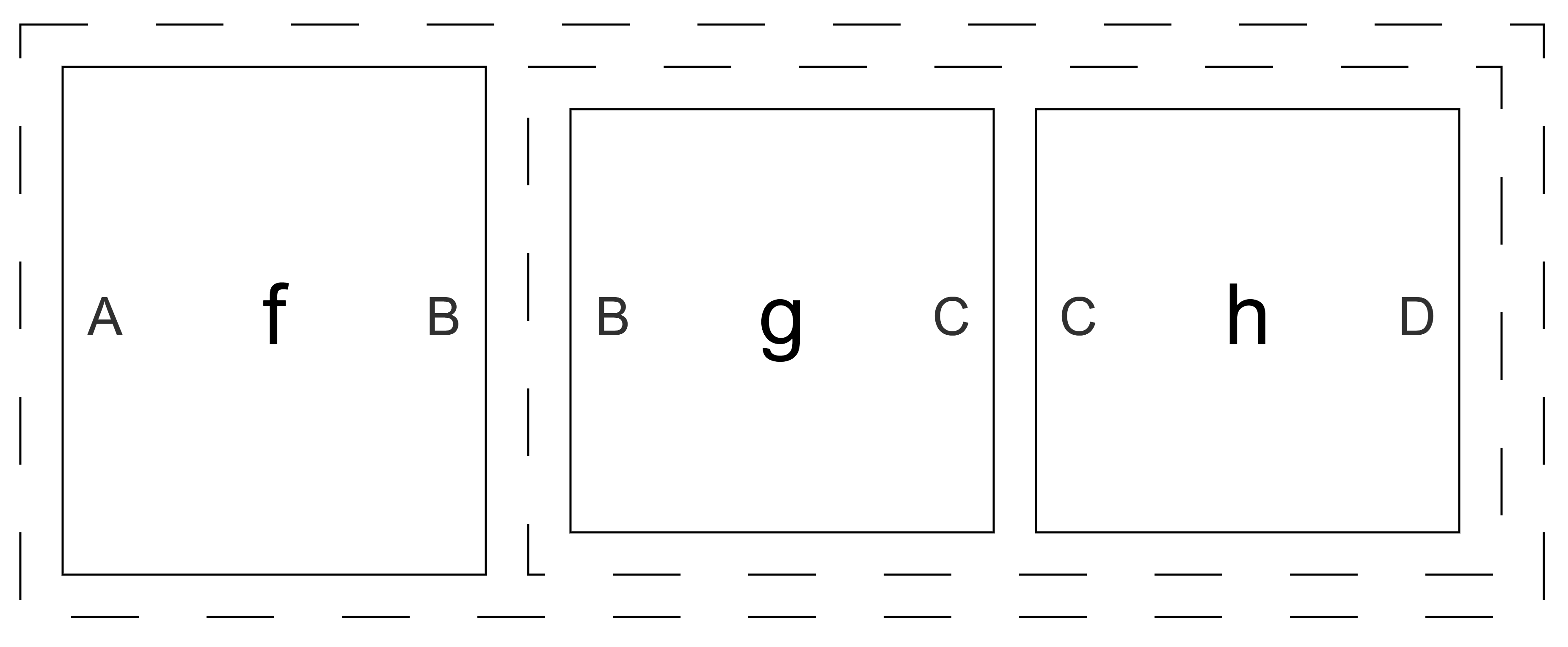}
  \caption{\vicar visualization of \coqe{f ∘ (g ∘ h)}.}
\end{subfigure}
\caption{How \vicar renders parentheses.}
\label{fig:assoc-example-b}
\end{figure}

Traditional string diagrams omit details such as associativity, so both $f \circ (g \circ h)$ and $(f \circ g) \circ h$ are visualized as~\Cref{fig:assoc-example-a}, for example. This is useful for pen-and-paper proofs but not for proof assistants---formal proofs require each ``obvious'' detail to be addressed. In Coq, $f \circ (g \circ h)$ and $(f \circ g) \circ h$ are two distinct terms, equivalent only via rewriting (applying a theorem which states they are equal). Identifying necessary rewrites is essential to the proof engineering workflow, so we tweaked \vicar's string diagram notation to maintain all parenthesizing explicitly. We represent these parentheses by circumscribing boxes, and the differences can be easily identified as in~\Cref{fig:assoc-example-b}. Color can optionally be added for clarity.

Our visualizations, while convenient for formal proof, are more verbose and add layers of complexity over the simpler diagrams. 
Future extensions of \vicar hope to solve this problem, by automatically handling structure.
This would allow for diagrams to be canonically represented and for proof engineers to no longer focus on structural rewrites.
While work has been done on rewriting in Coq modulo associativity and commutativity, we found none of it sufficient for what \vicar needs~\cite{braibant2011tactics}. The current gap between existing modulo associativity rules and this project relates to rules that depend on the interactions between two operators, like $(f \circ g) \otimes (h \circ t) \simeq (f \otimes h) \circ (g \otimes t)$. There are promising directions being developed on top of e-graph based equality saturation, as we discuss in \Cref{sec:future}.

\paragraph{Visualization semantics and workflow}

\begin{figure}
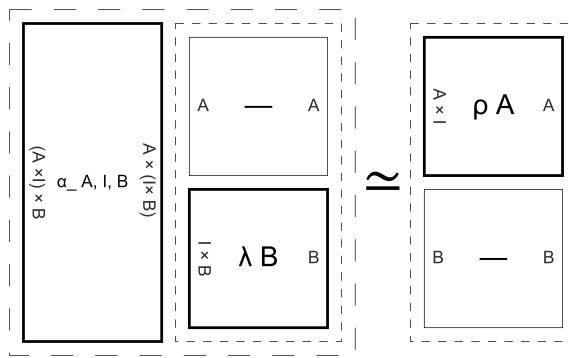

    \centering
    \pngfig{triangle}{0.5\linewidth}
    \caption{A visualization of the triangle identity.}
    \label{fig:triangle}
\end{figure}

\vicar allows for visualization of morphisms and morphism equivalences over base category instances up to symmetric monoidal ones. We unlock functionality as we deal with more expressive categories.
\begin{itemize}
    \item A morphism $f:A\leadsto B$ is visualized as a quadrilateral, marked with $A$ on the left and $B$ on the right. 
    \item The identity morphism for $A$ is denoted by a wire (horizontal line) with $A$ annotating the input and output positions. 
    \item Morphisms are composed by placing them side-by-side horizontally. 
    \item In monoidal categories, morphisms are tensored by placing them side-by-side vertically. 
    \item Taking the inverse of the morphism is rendered in a box attached to left of the morphism. 
    \item An isomorphism is a morphism with an emphasized bounding box.
    \item Category-specific terms, such as the associator, are identified as such (with the notation from \Cref{sec:Background}).
    \item In braided monoidal categories, the braiding is rendered as a big cross. 
\end{itemize} 
An example which uses several of these features can be seen in \Cref{fig:triangle}, which displays the statement of the triangle identity from \Cref{fig:coherence}.

\begin{figure}[ht]
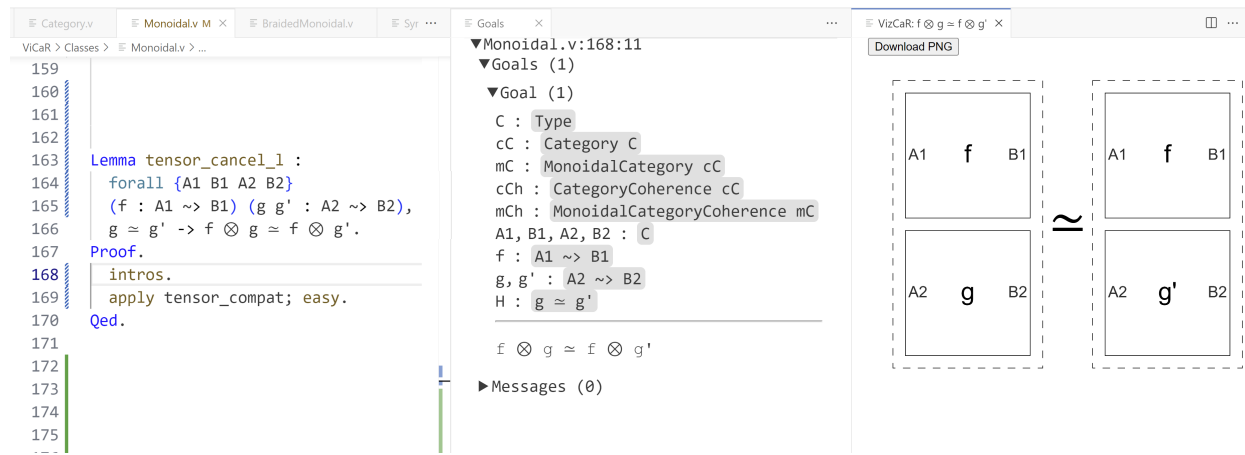

    \centering
    \pngfig{ide}{\linewidth}
    \caption{User's proof-writing IDE state}
    \label{fig:ide-state}
\end{figure}

One of \vicar's most convenient features is its integration into the Coq proof-writing workflow. \vicar is connected to the coq-lsp VSCode extension~\cite{lsp}, which type checks dynamically and prints the current goal state (hypotheses and statements yet to be proved) based on cursor position within a proof. The visualizer renders a string-diagrammatic goal state, alongside the printed one, and updates automatically as the goal changes. The user's setup can be seen in \Cref{fig:ide-state}. 

Because of this integration, the visualizer is able to use real-time hypothesis information to inform rendering choices. This is useful for distinguishing overloaded notation or opaque variables. For example, if the variable \coqe{f} in a proof state is a morphism from \coqe{A} to \coqe{B}, the visualizer will make this clear while the goal may not show it explicitly. Association and the effect of braiding may also be unclear from the proof state. The visualizer consults the hypothesis to check the type of each morphism and uses this information to label the inputs and outputs of each morphism's box.

\section{Automation}\label{sec:automation}
Categorical structure gives a well-defined domain on which we can develop partial proof automation.
Specifically, we can automate proofs of statements that two morphisms are equivalent. 
This is the form of many theorems in verification projects with monoidal structure. 
For example, theorems about matrices take the form that certain matrices are equal, and theorems about the ZX-calculus take the form that diagrams are proportional.
These types of proofs generally consist of rewrites and share common techniques, such as manipulation of the underlying categorical structure. 
In this section, we explain \vicar's tactics and how they can help with these proofs.

\paragraph{Coercing to categories} 
Coq will often fail to apply general categorical lemmas to proofs about specific typeclass instances because it cannot recognize which terms correspond to categorical structures. 
For example, suppose the category of matrices (defined in full in \Cref{subsec:matrices}) has been declared as a typeclass instance, with composition and identity given by matrix multiplication and the identity matrix. 
It would not be possible to rewrite the term \coqe{I_n $\times$ A} with the lemma whose statement is \coqe{id_ A $\circ$ f $\simeq$ f}. 
Coq fails to recognize that \coqe{$\times$} is the composition of a declared category instance and \coqe{I_n} is the identity.

To address this gap, we provide the tactic \coqe{categorify}. 
It performs setup necessary for our automation to identify the structure of the goal.
For the category of matrices, calling \coqe{categorify} would replace each instance \coqe{A $\times$ B} of matrix multiplication with categorical composition, \coqe{A $\circ$ B}, and the same for all other terms used to define to the instance.
In practice, when using \vicar, this serves as a setup tactic. At the beginning of the proof, \coqe{categorify} is called, which unlocks all the functionality of the visualizer and automation, provided the relevant structural and coherence instances have been defined. 
This tactic is purely aesthetic, in that it does not change the underlying goal, only the way in which that goal is presented. 
This change in presentation allows tactics to recognize categorical structures that are implicit. 

\paragraph{Foliation}

\begin{figure}
\centering
\begin{subfigure}{.25\textwidth}
    \centering
    \includegraphics[width=0.95\linewidth]{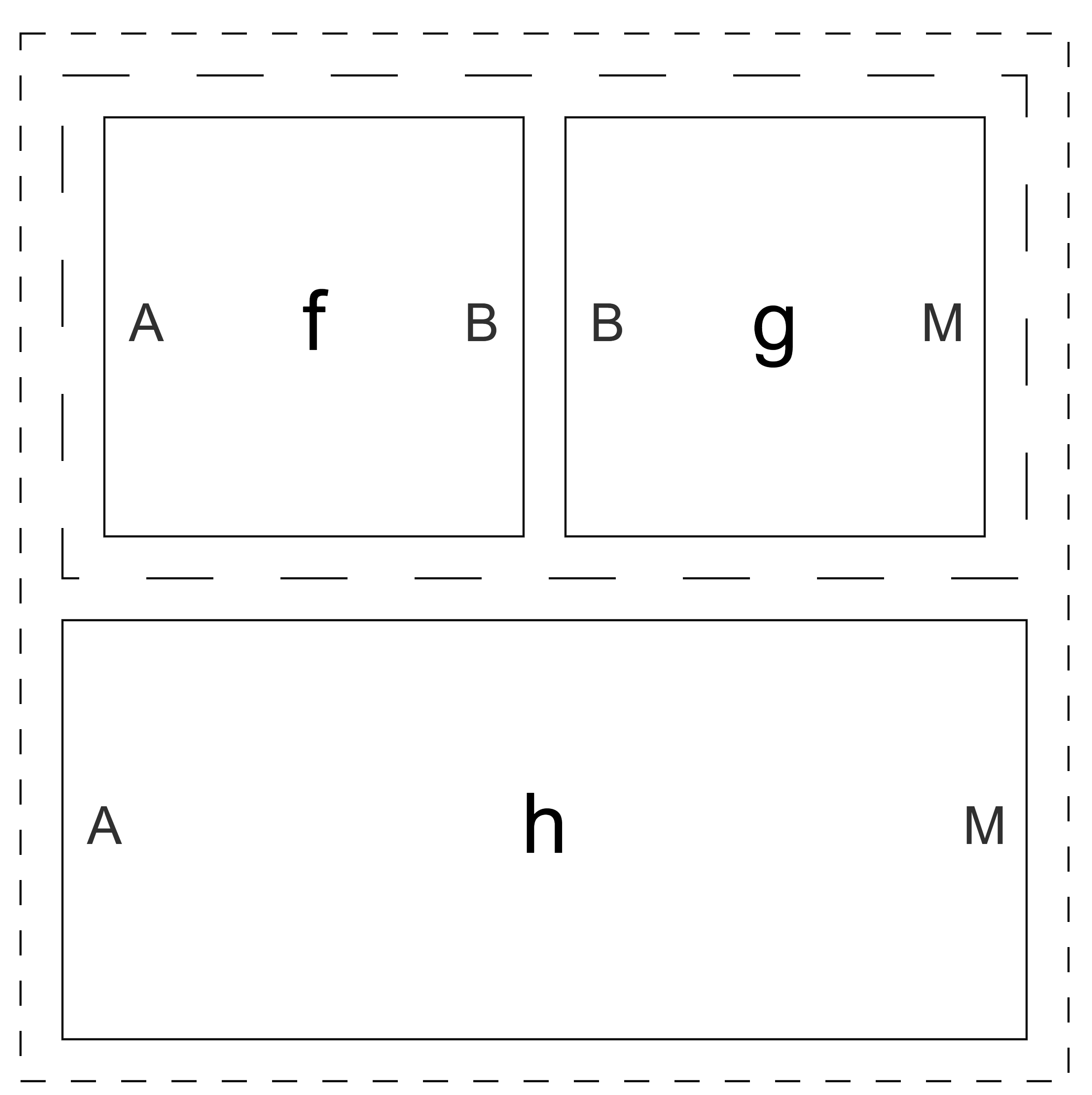}
    \caption{Initial state.}
\end{subfigure}
\begin{subfigure}{.27\textwidth}
  \centering
  \includegraphics[width=0.95\linewidth]{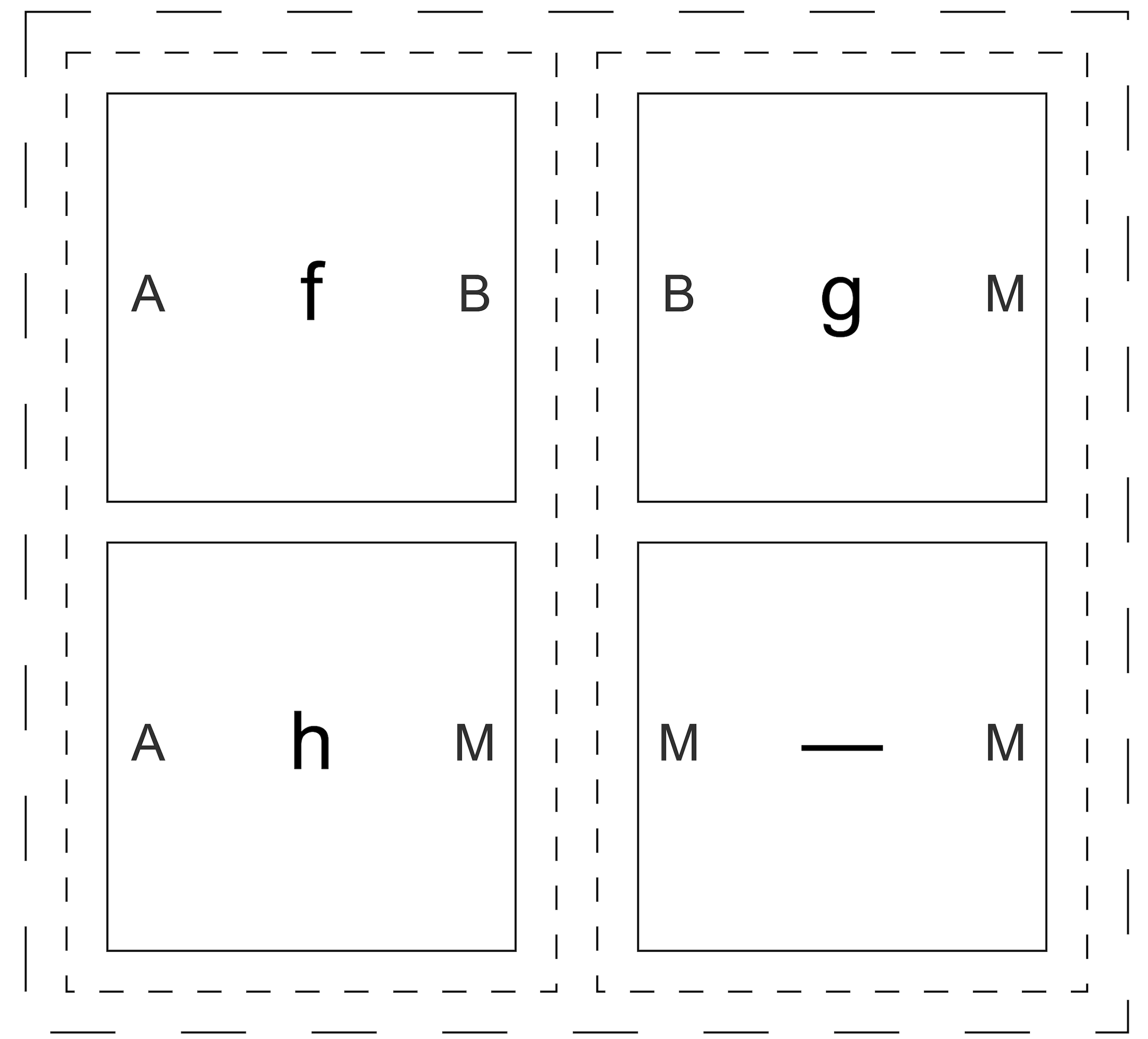}
  \caption{Weak foliation.}
\end{subfigure}
\begin{subfigure}{.375\textwidth}
  \centering
  \includegraphics[width=0.95\linewidth]{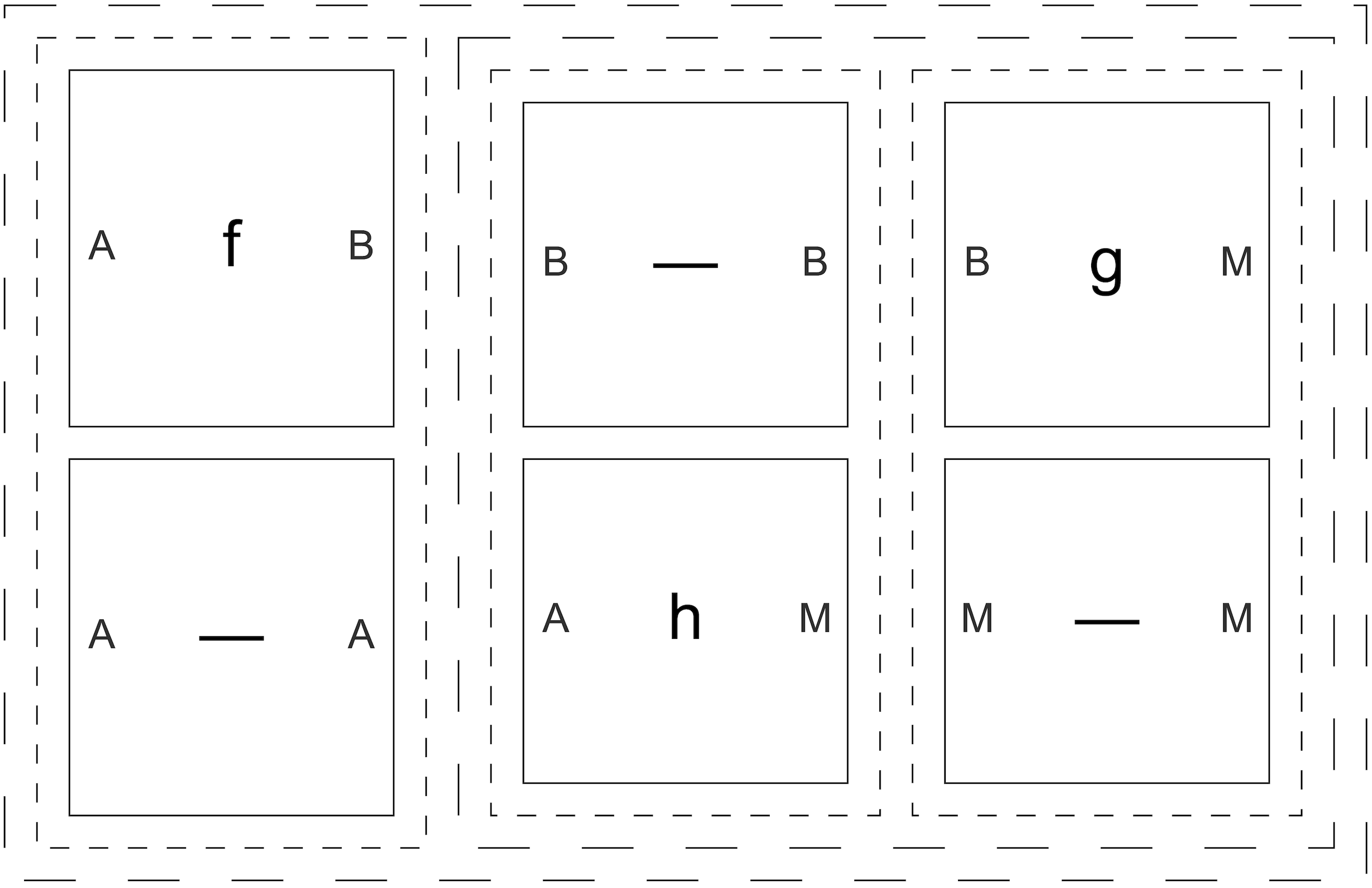}
  \caption{Foliation.}
\end{subfigure}
\caption{Visualization of foliation tactics on \coqe{(f ∘ g) ⊗ h}.}
\label{fig:foliation}
\end{figure}

A common representation of a morphism is a foliation~\cite{coecke2017picturing}. 
A foliation is a composition of ``stacks,'' each of which is the tensor of identity morphisms and a single non-identity morphism. 
Such a representation exists for any diagram and can be considered a standard form and useful for proofs~\cite{coecke2017picturing}. Formally proving this result for categorical instances, however, requires the complex notion that some morphisms are atomic with respect to this decomposition. For example, in the category of matrices, it is hard to define when a matrix should be decomposable with respect to matrix multiplication and Kronecker product (the tensor product). Our tactic \coqe{foliate} computes a foliation of a given morphism from any monoidal categorical instance, proves they are equivalent, and replaces the morphism with its foliation. 
Often, a full foliation is undesirable, and a more concise form is preferable. The \coqe{weak_foliate} tactic performs a partial foliation that allows multiple non-identity morphisms in a stack, but still ensures no stack contains a composition. 
An example of \coqe{weak_foliate} and \coqe{foliate} is given in \Cref{fig:foliation} on \coqe{(f ∘ g) ⊗ h}. Its partial foliation is \coqe{f ⊗ h ∘ g ⊗ id_ M} and its full foliation is \coqe{f ⊗ id_ A ∘ (id_ B ⊗ h ∘ g ⊗ id_ M)}.

\paragraph{Associativity: partnering and rewriting}

A recurring task in many Coq projects is dealing with associativity. Coq expressions have explicit association, so to rewrite a subterm of a sequence of compositions, it is often necessary to perform several rewrites using associativity rules. Even the expression \coqe{g $\circ$ f $\circ$ f^-1}
cannot be rewritten to \coqe{g} directly, requiring first a rewrite using the associativity condition. In larger expressions, the reassociations which must be performed can be quite laborious to specify manually. We provide several tactics to automate this work.

First, we define the \coqe{partner} tactic, which takes two terms as arguments and attempts to reassociate the goal to make these terms syntactically adjacent. After calling \coqe{partner f g}, we get \coqe{(f ∘ g)} as a subterm of the goal, assuming it can isolated by reassociation.

Building on this technique, we also define the \coqe{assoc_rw} tactic, which takes a lemma as an argument and attempts to reassociate the goal such that the lemma can be rewritten. This works for any lemma whose conclusion, possibly quantified over arguments, has the form $F\simeq g$, where $F$ is some sequence of compositions and $g$ is any morphism. This tactic allows a user to entirely ignore the association of the goal, and simply rewrite according to the sequence of morphisms present in the goal. 
For example, given a lemma \coqe{f $\circ$ g $\simeq$ h}, \coqe{assoc_rw} would rewrite its occurrence in the expression \coqe{i $\otimes$ (e $\circ$ f $\circ$ g)}, even though this requires reassociation within the argument to the tensor product.

These tactics together entirely obviate the need for manual association of the goal. This delivers on one part of the motivation for using string diagram representations of morphisms: such representations implicitly encode the coherence conditions of monoidal categories by means of topological irrelevance~\cite{Selinger2010}. While full topological irrelevance is hard both to define and to prove within Coq, suppressing the consideration of association is a first step in this direction.

\paragraph{Simplifying the goal state}
\begin{lstlisting}[caption=\vicar's tactics being used to easily solve an example lemma.,label=fig:tactic,
float=tp,floatplacement=tbp %
]
Lemma assoc_rw_example {A B M N : Cobj}
  (f : A mor B) (g : M mor N) : 
  (β_ A, M)^-1 ∘ f ⊗ g ∘ β_ B, N
  ≃ g ⊗ f.
Proof. 
  assoc_rw braiding_natural. (* apply braiding lemma *)
  cancel_isos. (* cancellation *)
  reflexivity.
Qed.
\end{lstlisting}

\begin{figure}
\begin{subfigure}{.42\textwidth}
  \centering
  \includegraphics[width=0.95\linewidth]{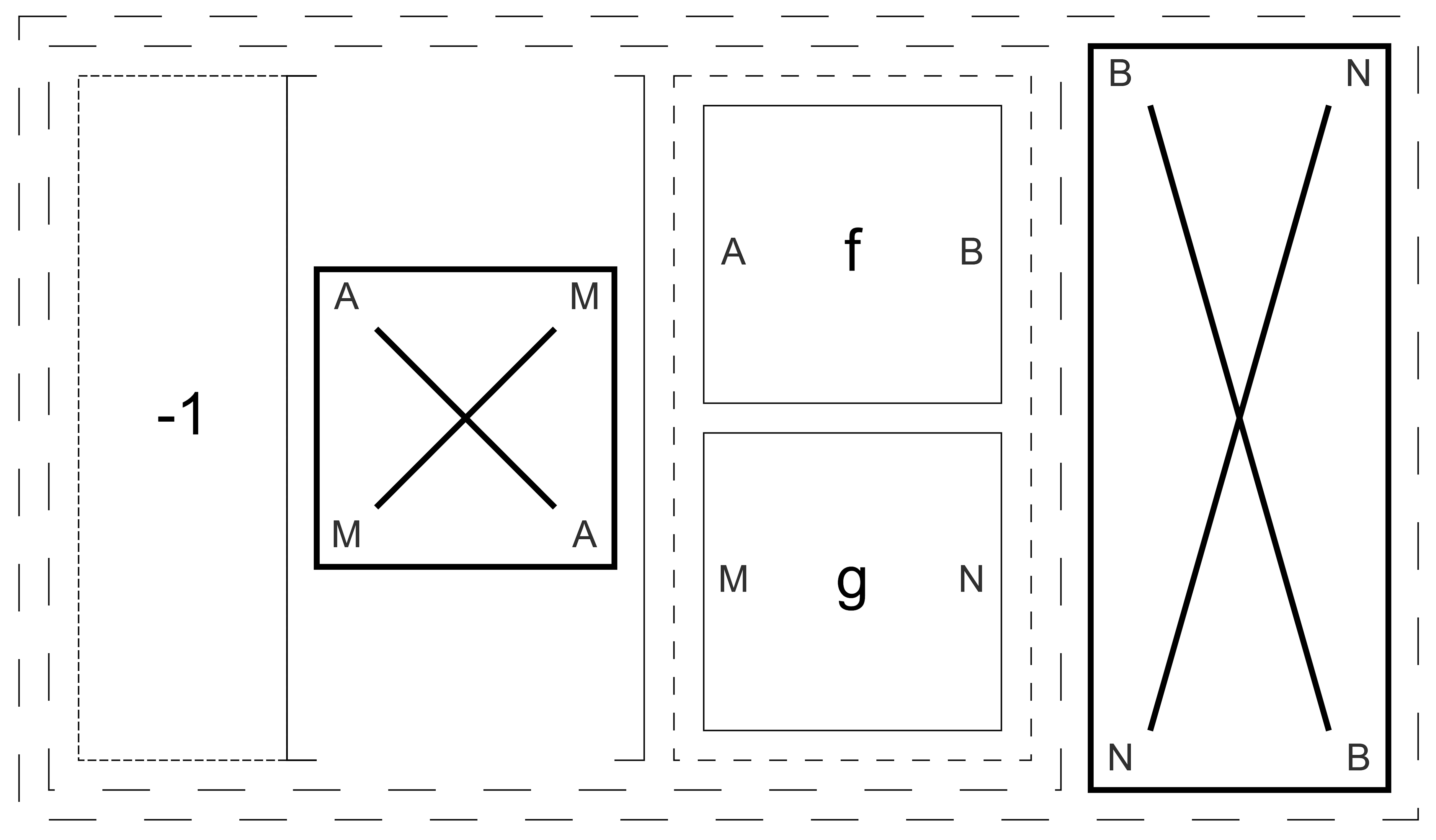}
  \caption{Initial goal state.}
\end{subfigure}
\begin{subfigure}{.42\textwidth}
  \centering
  \includegraphics[width=0.95\linewidth]{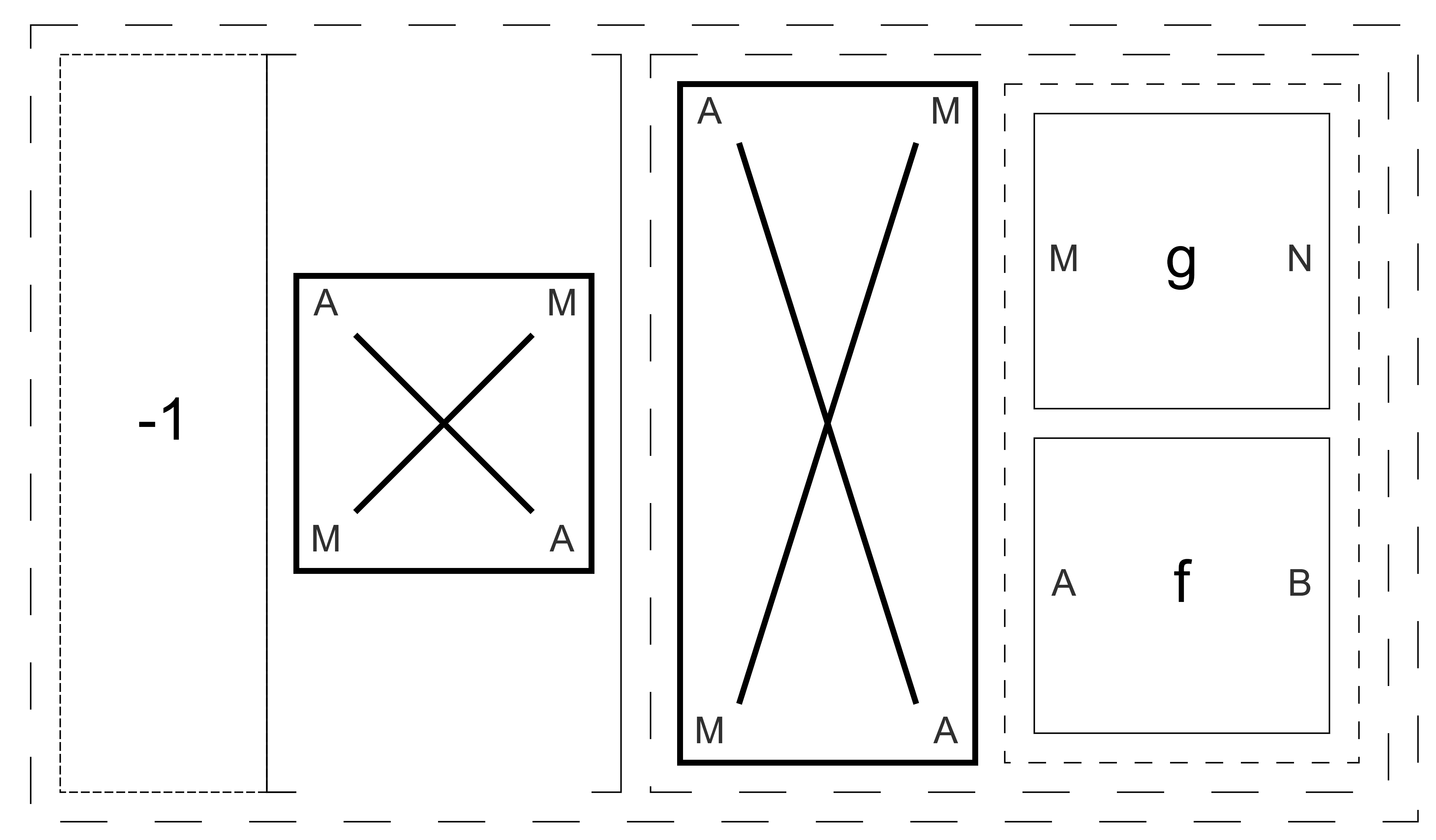}
  \caption{Applying a braiding lemma.}
\end{subfigure}
\begin{subfigure}{.132\textwidth}
  \centering
  \includegraphics[width=0.95\linewidth]{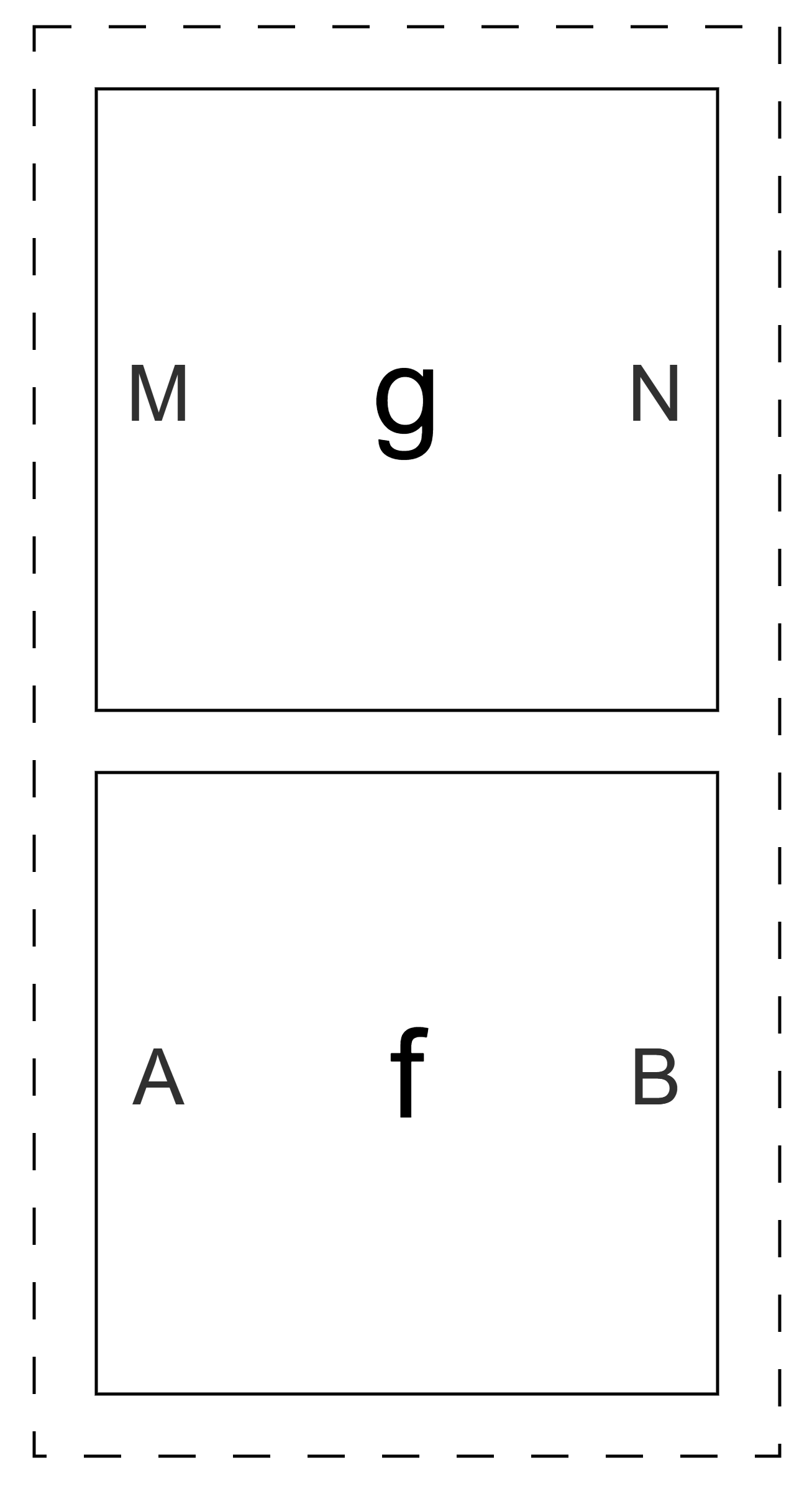}
  \caption{Canceling.}
\end{subfigure}
\caption{Visualization at each step of \Cref{fig:tactic}.}
\label{fig:tactics-diagrams}
\end{figure}

Building on these tactics, we provide a number of tactics to simplify the goal state. The tactic \coqe{cancel_isos} will cancel any isomorphism adjacent to its inverse, independent of associativity. This is particularly useful 
for eliminating structural morphisms that accumulate by applications of naturality properties. The tactic \coqe{cat_simpl} combines this cancellation with the removal of identity morphisms. These tactics give a quick way to perform common simplifications of a goal without having to consider association.
The tactic \coqe{cat_easy} attempts to trivial goals that are fundamentally structural. It will  cancel isomorphisms and identities, right-associate the goal, and perform weak foliation until the goal is solvable by reflexivity.

\Cref{fig:tactic} and \Cref{fig:tactics-diagrams} give an example of these tactics in action to almost completely automate a proof. The \coqe{assoc_rw} tactic automatically associates the goal to apply the given lemma, a braiding coherence condition. \coqe{cancel_isos} then reassociates and cancels out the braiding and its inverse. These tactics can be applied to any proof for a concrete instance, after first calling \coqe{categorify}.

\paragraph{Monoidal coherence}
We formalize the coherence theorem for monoidal categories\footnote{Our proof follows exactly the method of \cite{beylin1996coherence}, to which we refer for details. As in their proof, our result applies only to types with the property of Uniqueness of Identity Proofs. It is not obvious how to overcome this requirement for this proof or for the proof of MacLane, but in practical cases, such as for natural numbers, this property is provable. Moreover, this property is immediate in classical logic, and in particular for types with decidable equality.} in \vicar and use it to create a tactic, \coqe{monoidal}, for solving morphism equivalences that follow directly from monoidal structure. It therefore implements an approximation to traditional string diagram reasoning. 
This tactic uses proof by reflection, meaning it transforms the goal into an equivalent one provable by pure computation. 

The tactic first converts both sides of the goal into inductively-defined intermediate representations. It then applies a theorem that states that two such inductive diagrams are equivalent if their images under a certain function are the same. That function converts the diagrams to a permissive form which mimics traditional string diagrams, meaning without any information about associativity of composition or the tensor product. Finally, the function performs simplification by removing structural isomorphisms, sliding non-identity morphisms as far left through identity morphisms as possible, then removing any extraneous identity morphisms. If the resulting forms are identical, the tactic proves the original morphisms are equivalent. \Cref{fig:monoidal-tactic} demonstrates the ability of \coqe{monoidal} to easily solve certain goals.

\begin{figure}
  \centering
  \includegraphics[width=0.95\linewidth]{monoidal-ex.png}
  \caption{A morphism equivalence lemma which can be immediately proved with one invocation of \coqe{monoidal}.}
\label{fig:monoidal-tactic}
\end{figure}

\section{Example uses of \vicar}\label{sec:examples}

\subsection{\vyzx}

The \zxcalc is a graphical language to represent quantum operations~\cite{coecke-duncan-zx}. 
Each \zxdiag represents a linear transformation between qubits and consists of red and green nodes connected by wires. 
ZX's appeal is that its rewrite rules are easily visualized as graph manipulations. 
\zxdiags can be visually transformed into any equivalent diagram using a finite set of rewrites. 
The \zxcalc has been used for simulation~\cite{kissinger2022simulating}, circuit optimization~\cite{pyzx}, and fault tolerance~\cite{bombin2023unifying} work.
For a deeper introduction to the \zxcalc, refer to one of these surveys~\cite{vandewetering2020zxcalculus,coecke2023basic}.

\vyzx (Verifying the ZX-Calculus) is an effort to formalize the ZX-calculus in Coq~\cite{2023vyzx}.
It gives inductive definitions for diagrams and interprets them through standard matrix semantics for the \zxcalc. 
The \zxcalc corresponds to a symmetric monoidal category whose objects are natural numbers (representing the number of input and output wires) and whose morphisms are \zxdiags. Horizontal composition of morphisms (connecting the input and output wires of two diagrams) is associative.
The tensor product is given by vertically stacking diagrams, corresponding to addition on natural numbers with identity object $0$.
The morphism identity given for a natural number $n$ is $n$ wires with no nodes, since composition with this object does not affect any diagram.

As \vyzx is a well-developed verification library, most of these categorical structures and necessary lemmas already existed. 
As a result, the coherence conditions for these morphisms were straightforward to prove. 
The braiding was easily implemented, but proving naturality was a significant task. 
Fortunately, this task was made easier by proving naturality of braiding for matrices, as discussed in \ref{subsec:matrices}. 

The VyZX visualizer, \vizx, displays ZX-diagrams by parsing an expression into its building blocks and generating a string diagram. We modified \vizx to work for general category instances, abstracting the specific display style used for \zxdiags and significantly reworking the foundation of the visualizer to work in the context of a general category. \Cref{fig:zx-example} gives an example of \vicar's visualizer for a ZX-specific lemma, where \coqe{zx0}, \coqe{zx1}, and \coqe{zx2} are \zxdiags. Its string diagram makes clear this lemma is really just a structural result.

\vicar grew out of \vyzx in an attempt to separate rules specific to the \zxcalc from rules that are common to any symmetric monoidal category.
Typical \zxcalc literature does not reason about such structure but rather about \zxdiags as graphs.
Therefore, future \vyzx development can benefit from \vicar's automation by being able to better ignore structural manipulation of \zxdiags.

\begin{figure}
    \centering
    \pngfig{zx-viz}{0.6\linewidth}
    \caption{Visualization of the statement 
    \coqe{zx0 ⟷ n_wire m ↕ (n_wire o ⟷ zx1) ⟷ zx2 ∝ 
    zx0 ↕ zx1 ⟷ zx2.}}
    \label{fig:zx-example}
\end{figure}

\subsection{Calculus of Relations}

\begin{figure}
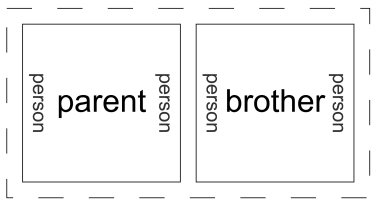

    \centering
    \pngfig{uncle-relation}{0.3\linewidth}
    \caption{Visualizing the ``uncle'' relation between two people, \coqe{parent ∘ brother}. (Note that this is diagrammatic composition; parent precedes brother.)}
    \label{fig:uncle-example}
\end{figure}

In contrast to the previous implementation, the calculus of relations is not a pre-existing Coq project. Its implementation follows the book \emph{Picturing Quantum Processes}~\cite{coecke2017picturing}. To adapt this work to Coq, we define the objects of our category to be types, and we define morphisms between types \coqe{T} and \coqe{S} to have type \coqe{T -> S -> Prop}. As this is defined over any \coqe{T} and \coqe{S}, it can be easily applied to any specific example. By defining base relations such as \coqe{sibling} and \coqe{parent} over a type \coqe{person}, we can construct and visualize relations such as \coqe{uncle} in \Cref{fig:uncle-example}. While relations are a toy example, they do demonstrate some insights.

Our implementation of relations was made with \vicar in mind, and so all proofs are intended for our typeclasses. However, categorical structure is far from a complete description of the properties of relations. For example, it is natural to want to describe the transitive closure of a relation \coqe{R : T -> T -> Prop}. There is no clear way to do this solely using categorical construction.
Instead, we are able to use Coq's built-in capabilities to describe a very similar construct: given \coqe{R : T -> T -> Prop}, we define \coqe{R^n}, the repeated application of \coqe{R} $n$ times. 
We can then construct the transitive closure diagrammatically as \coqe{$\rho$_ A $\circ$ (id_ A $\otimes$ any_nat) $\circ$ R^n}, where \texttt{any\_nat} is a relation that relates \coqe{unit} to any \coqe{nat}. 
By having \vicar embedded in an existing proof assistant, we gain the ability to easily create new morphisms as needed and can reason about both structure and properties specific to a particular instance, together.

\subsection{Matrices}
\label{subsec:matrices}

The collection of vector spaces over a field $k$ form the category $\textrm{Vect}_k$ whose morphisms are linear transformations of spaces. This is a monoidal category over the tensor product of vector spaces whose identity element is $k$. Moreover, this monoidal category is symmetric, as $V \otimes W \cong W \otimes V$.

A commonly used subcategory of $\textrm{Vect}_k$ is the category $\textrm{FinDimVect}_k$ of finite-dimensional vector spaces, which is a symmetric monoidal category in exactly the same way. $\textrm{FinDimVect}_k$ has a skeleton given by vector spaces of the form $k^n$ for $n\in \mathbb{N}$. This skeleton is isomorphic to the category whose objects are natural numbers and whose morphisms from $n$ to $m$ are the $n\times m$ matrices over $k$, representing linear transformations $k^n\rightarrow k^m$. We implement that category for $k=\mathbb{C}$ using the matrices in the \qlib library~\cite{QuantumLib}, an existing Coq library of verified mathematics for quantum computing. \qlib is used as the basis of several quantum computing-related projects, including \vyzx, \sqir, \voqc, and \qwire \cite{2023vyzx, hietala2021sqir, hietala2021voqc, paykin2017qwire}.

\qlib already implements most of the definitions and lemmas required to instantiate a category. Matrices in \qlib are defined as functions $\N \rightarrow \N \rightarrow \mathbb{C}$. The dimensions of \qlib matrices are not enforced by the type checker, but instead given as ``phantom types'' to guide the programmer~\cite{rand2018phantom}. This lack of strict dimensions makes the construction and use of the associator and unitors very straightforward. 

We implement our category instances using bounded equivalence ($\equiv$) as our morphism equivalence, which says that matrices $A,B$ of dimension $n\times m$ are equivalent if they are equal on all entries within their bounds.
The majority of the work in showing matrices are a braided monoidal category involves showing the naturality condition of the braiding. In this category, the braiding is given by the commutation matrices $K_{n,m}$, which have the property that for any $n\times m$ matrix $A$ and $p\times q$ matrix $B$, we have
 \(
 K_{p,n} (A \otimes B) = (B \otimes A) K_{q,m}
 \)~\cite{magnus79}. 
 Due to phantom types, the associator and unitor can be defined as the identity matrix interpreted with the appropriate dimensions. In proofs, these matrices can be immediately canceled, making the coherence conditions easy to prove.

A diagrammatic representation of matrix expressions involving the Kronecker encodes facts that textual representations do not. For example, \Cref{fig:matrix-ex} shows the distributivity of the Kronecker product over matrix multiplication. The diagram highlights the structure of the vector spaces on which these matrices act, making a non-obvious equation more believable. 
\begin{figure}
    \centering
    \pngfig{mx-viz}{0.6\linewidth}
    \caption{Visualization of the Kronecker mixed-product, $(A \otimes C) \times (B \otimes D) \equiv (A \times B) \otimes (C \times D)$. The variables $B$ and $C$ appear swapped in the text, but the diagram reveals that only the associativity changes.}
    \label{fig:matrix-ex}
    \vspace{-1.0em}
\end{figure}

\section{Related Work}

Category theory has been formalized many times in Coq. Some formalizations such as Coq-HoTT~\cite{gross2014experience} and UniMath~\cite{UniMath} are built on top of homotopy type theory (HoTT)~\cite{hottbook}. Other works such as Coq-Categories~\cite{megacz2007coinductive} built category theory using typeclasses as we did. However, a commonality between all the projects we observed is that their goal is to verify large category-theoretic results, and different libraries focus on different parts of category theory~\cite{gross2014experience}. Ultimately, we choose to create our own library to minimize dependencies, such as on HoTT, and allow ourselves flexibility to change parts of the library as needed, since we uniquely focus on automation and applications.

Category theory leans on visual intuition, so it should be no surprise that visualization has played a central role in formalizing categories. Many visualizers display commutative diagrams, rather than string diagrams, and several of these projects have robust interactive editors, such as YADE~\cite{lafont2024diagram}. Other projects such as Chyp~\cite{chyp} focus on specifying theories and visualizing abstract string diagrams, rather than being applicable to other projects. Another visualizer, commutative-diagrams~\cite{chabassier2023graphical}, adds tactics and interaction for manipulating string diagrams in Coq, though it does not use associativity information as we do. Our visualizer lacks interaction, but all the interactive visualizers we found have some dependency on UniMath, which replaces some of Coq's core libraries and can cause compatibility issues.

\section{Future Directions}\label{sec:future}

\vicar provides a framework for visualization and automated rewriting by defining typeclasses for categorical and monoidal structure. 
The language of string diagrams can further describe rigid symmetric monoidal categories, with the unit and counit of the category depicted as half turns~\cite{Selinger2010}. 
\vicar can be extended with this structure, allowing it to capture a wide variety of process theories.

Many symmetric monoidal categories are interpreted using a semantics function.
In fact, the semantics are often used to define morphism equivalence.
We aim to provide infrastructure within \vicar to allow translation between different symmetric monoidal categories for semantic interpretation.
This would greatly simplify constructs in categories equipped with a \vicar-based semantics by lifting them.

Seeing the advantage that the \coqe{assoc_rw} tactic gives us, we want to further expand the capabilities of rewriting structure.
In fact, we would like to completely be able to rewrite without worrying about structure.
To achieve this goal, we are working on an e-graph equality saturation-based solver for different structural configurations.
The solver will work by ingesting Coq statements into a custom AST, and then the e-graph solver will use the structural rewrite rules in \vicar.
Once a proof is found, it will be exported into Coq and checked.
To build this proof, one must translate dependent type information into an e-graph system.
We propose handling by translating SMC structure without dependent types and then rebuilding dependent types as functions within the e-graph.
With this system, we will have all the guarantees we need, as rules can check the dependent type reconstruction function output, given that the diagram given to the solver is valid. This will further facilitate practical reasoning about categorical structures in Coq.

\section*{Acknowledgments}
This material is based upon work supported by the Air Force Office of Scientific Research under award numbers FA95502310361 and FA95502310406, and
EPiQC, an NSF Expedition in Computing, under Grant No. CCF-1730449.

\bibliography{references}

\begin{thebibliography}{10}
\providecommand{\bibitemdeclare}[2]{}
\providecommand{\surnamestart}{}
\providecommand{\surnameend}{}
\providecommand{\urlprefix}{Available at }
\providecommand{\url}[1]{\texttt{#1}}
\providecommand{\href}[2]{\texttt{#2}}
\providecommand{\urlalt}[2]{\href{#1}{#2}}
\providecommand{\doi}[1]{doi:\urlalt{https://doi.org/#1}{#1}}
\providecommand{\eprint}[1]{arXiv:\urlalt{https://arxiv.org/abs/#1}{#1}}
\providecommand{\bibinfo}[2]{#2}

\bibitemdeclare{inproceedings}{beylin1996coherence}
\bibitem{beylin1996coherence}
\bibinfo{author}{Ilya \surnamestart Beylin\surnameend} \&
  \bibinfo{author}{Peter \surnamestart Dybjer\surnameend}
  (\bibinfo{year}{1996}): \emph{\bibinfo{title}{Extracting a proof of coherence
  for monoidal categories from a proof of normalization for monoids}}.
\newblock In \bibinfo{editor}{Stefano \surnamestart Berardi\surnameend} \&
  \bibinfo{editor}{Mario \surnamestart Coppo\surnameend}, editors: {\slshape
  \bibinfo{booktitle}{Types for Proofs and Programs}},
  \bibinfo{publisher}{Springer Berlin Heidelberg}, \bibinfo{address}{Berlin,
  Heidelberg}, pp. \bibinfo{pages}{47--61}, \doi{10.1007/3-540-61780-9\_61}.

\bibitemdeclare{misc}{bombin2023unifying}
\bibitem{bombin2023unifying}
\bibinfo{author}{Hector \surnamestart Bombin\surnameend},
  \bibinfo{author}{Daniel \surnamestart Litinski\surnameend},
  \bibinfo{author}{Naomi \surnamestart Nickerson\surnameend},
  \bibinfo{author}{Fernando \surnamestart Pastawski\surnameend} \&
  \bibinfo{author}{Sam \surnamestart Roberts\surnameend}
  (\bibinfo{year}{2023}): \emph{\bibinfo{title}{Unifying flavors of fault
  tolerance with the ZX calculus}}, \doi{10.22331/q-2024-06-18-1379}.
\newblock \eprint{2303.08829}.

\bibitemdeclare{inproceedings}{braibant2011tactics}
\bibitem{braibant2011tactics}
\bibinfo{author}{Thomas \surnamestart Braibant\surnameend} \&
  \bibinfo{author}{Damien \surnamestart Pous\surnameend}
  (\bibinfo{year}{2011}): \emph{\bibinfo{title}{Tactics for reasoning modulo AC
  in Coq}}.
\newblock In: {\slshape \bibinfo{booktitle}{International Conference on
  Certified Programs and Proofs}}, \bibinfo{organization}{Springer}, pp.
  \bibinfo{pages}{167--182}, \doi{10.1007/978-3-642-25379-9\_14}.

\bibitemdeclare{misc}{castello2023inductive}
\bibitem{castello2023inductive}
\bibinfo{author}{Jonathan \surnamestart Castello\surnameend},
  \bibinfo{author}{Patrick \surnamestart Redmond\surnameend} \&
  \bibinfo{author}{Lindsey \surnamestart Kuper\surnameend}
  (\bibinfo{year}{2023}): \emph{\bibinfo{title}{Inductive diagrams for causal
  reasoning}}.
\newblock \eprint{2307.10484}.

\bibitemdeclare{inproceedings}{chabassier2023graphical}
\bibitem{chabassier2023graphical}
\bibinfo{author}{Luc \surnamestart Chabassier\surnameend} \&
  \bibinfo{author}{Bruno \surnamestart Barras\surnameend}
  (\bibinfo{year}{2023}): \emph{\bibinfo{title}{A graphical interface for
  diagrammatic proofs in proof assistants}}.
\newblock In: {\slshape \bibinfo{booktitle}{29th International Conference on
  Types for Proofs and Programs TYPES 2023--Abstracts}},
  p.~\bibinfo{pages}{98}.

\bibitemdeclare{misc}{coecke2023basic}
\bibitem{coecke2023basic}
\bibinfo{author}{Bob \surnamestart Coecke\surnameend} (\bibinfo{year}{2023}):
  \emph{\bibinfo{title}{Basic ZX-calculus for students and professionals}}.
\newblock \eprint{2303.03163}.

\bibitemdeclare{inproceedings}{coecke-duncan-zx}
\bibitem{coecke-duncan-zx}
\bibinfo{author}{Bob \surnamestart Coecke\surnameend} \& \bibinfo{author}{Ross
  \surnamestart Duncan\surnameend} (\bibinfo{year}{2008}):
  \emph{\bibinfo{title}{Interacting Quantum Observables}}.
\newblock In \bibinfo{editor}{Luca \surnamestart Aceto\surnameend},
  \bibinfo{editor}{Ivan \surnamestart Damg{\aa}rd\surnameend},
  \bibinfo{editor}{Leslie~Ann \surnamestart Goldberg\surnameend},
  \bibinfo{editor}{Magn{\'u}s~M. \surnamestart Halld{\'o}rsson\surnameend},
  \bibinfo{editor}{Anna \surnamestart Ing{\'o}lfsd{\'o}ttir\surnameend} \&
  \bibinfo{editor}{Igor \surnamestart Walukiewicz\surnameend}, editors:
  {\slshape \bibinfo{booktitle}{Automata, Languages and Programming}},
  \bibinfo{publisher}{Springer Berlin Heidelberg}, \bibinfo{address}{Berlin,
  Heidelberg}, pp. \bibinfo{pages}{298--310},
  \doi{10.1007/978-3-540-70583-3\_25}.

\bibitemdeclare{book}{coecke2017picturing}
\bibitem{coecke2017picturing}
\bibinfo{author}{Bob \surnamestart Coecke\surnameend} \& \bibinfo{author}{Aleks
  \surnamestart Kissinger\surnameend} (\bibinfo{year}{2017}):
  \emph{\bibinfo{title}{Picturing Quantum Processes: A First Course in Quantum
  Theory and Diagrammatic Reasoning}}.
\newblock \bibinfo{publisher}{Cambridge University Press},
  \doi{10.1017/9781316219317}.

\bibitemdeclare{misc}{Coq12}
\bibitem{Coq12}
\bibinfo{author}{The \surnamestart {Coq} {Development}~{Team}\surnameend}
  (\bibinfo{year}{2012}): \emph{\bibinfo{title}{The {Coq} Reference Manual,
  version 8.4}}.
\newblock \bibinfo{note}{Available electronically at
  \url{http://coq.inria.fr/doc}}.

\bibitemdeclare{misc}{lsp}
\bibitem{lsp}
\bibinfo{author}{The Coq~LSP \surnamestart Developers\surnameend}
  (\bibinfo{year}{2023}): \emph{\bibinfo{title}{{G}it{H}ub - ejgallego/coq-lsp:
  {V}isual {S}tudio {C}ode {E}xtension and {L}anguage {S}erver {P}rotocol for
  {C}oq --- github.com}}.
\newblock \bibinfo{howpublished}{\url{https://github.com/ejgallego/coq-lsp}}.

\bibitemdeclare{inproceedings}{eilenberg1966closed}
\bibitem{eilenberg1966closed}
\bibinfo{author}{Samuel \surnamestart Eilenberg\surnameend} \&
  \bibinfo{author}{G~Max \surnamestart Kelly\surnameend}
  (\bibinfo{year}{1966}): \emph{\bibinfo{title}{Closed categories}}.
\newblock In: {\slshape \bibinfo{booktitle}{Proceedings of the Conference on
  Categorical Algebra: La Jolla 1965}}, \bibinfo{organization}{Springer}, pp.
  \bibinfo{pages}{421--562}, \doi{10.1007/978-3-642-99902-4\_22}.

\bibitemdeclare{inproceedings}{gross2014experience}
\bibitem{gross2014experience}
\bibinfo{author}{Jason \surnamestart Gross\surnameend}, \bibinfo{author}{Adam
  \surnamestart Chlipala\surnameend} \& \bibinfo{author}{David~I \surnamestart
  Spivak\surnameend} (\bibinfo{year}{2014}): \emph{\bibinfo{title}{Experience
  implementing a performant category-theory library in Coq}}.
\newblock In: {\slshape \bibinfo{booktitle}{Interactive Theorem Proving: 5th
  International Conference, ITP 2014, Held as Part of the Vienna Summer of
  Logic, VSL 2014, Vienna, Austria, July 14-17, 2014. Proceedings 5}},
  \bibinfo{organization}{Springer}, pp. \bibinfo{pages}{275--291},
  \doi{10.1007/978-3-319-08970-6\_18}.

\bibitemdeclare{inproceedings}{hietala2021sqir}
\bibitem{hietala2021sqir}
\bibinfo{author}{Kesha \surnamestart Hietala\surnameend},
  \bibinfo{author}{Robert \surnamestart Rand\surnameend},
  \bibinfo{author}{Shih-Han \surnamestart Hung\surnameend},
  \bibinfo{author}{Liyi \surnamestart Li\surnameend} \&
  \bibinfo{author}{Michael \surnamestart Hicks\surnameend}
  (\bibinfo{year}{2021}): \emph{\bibinfo{title}{{Proving Quantum Programs
  Correct}}}.
\newblock In \bibinfo{editor}{Liron \surnamestart Cohen\surnameend} \&
  \bibinfo{editor}{Cezary \surnamestart Kaliszyk\surnameend}, editors:
  {\slshape \bibinfo{booktitle}{12th International Conference on Interactive
  Theorem Proving (ITP 2021)}}, {\slshape \bibinfo{series}{Leibniz
  International Proceedings in Informatics (LIPIcs)}} \bibinfo{volume}{193},
  \bibinfo{publisher}{Schloss Dagstuhl -- Leibniz-Zentrum f{\"u}r Informatik},
  \bibinfo{address}{Dagstuhl, Germany}, pp. \bibinfo{pages}{21:1--21:19},
  \doi{10.4230/LIPIcs.ITP.2021.21}.

\bibitemdeclare{article}{hietala2021voqc}
\bibitem{hietala2021voqc}
\bibinfo{author}{Kesha \surnamestart Hietala\surnameend},
  \bibinfo{author}{Robert \surnamestart Rand\surnameend},
  \bibinfo{author}{Shih-Han \surnamestart Hung\surnameend},
  \bibinfo{author}{Xiaodi \surnamestart Wu\surnameend} \&
  \bibinfo{author}{Michael \surnamestart Hicks\surnameend}
  (\bibinfo{year}{2021}): \emph{\bibinfo{title}{A Verified Optimizer for
  Quantum Circuits}}.
\newblock {\slshape \bibinfo{journal}{Proc. ACM Program. Lang.}}
  \bibinfo{volume}{5}(\bibinfo{number}{POPL}), \doi{10.1145/3434318}.

\bibitemdeclare{misc}{QuantumLib}
\bibitem{QuantumLib}
\bibinfo{author}{\surnamestart {INQWIRE Developers}\surnameend}
  (\bibinfo{year}{2022}): \emph{\bibinfo{title}{{INQWIRE QuantumLib}}}.
\newblock \urlprefix\url{https://github.com/inQWIRE/QuantumLib}.

\bibitemdeclare{misc}{jwiegleycategory}
\bibitem{jwiegleycategory}
\bibinfo{author}{\surnamestart {John Wiegley}\surnameend}
  (\bibinfo{year}{2022}): \emph{\bibinfo{title}{{Category Theory in Coq}}}.
\newblock \urlprefix\url{https://github.com/jwiegley/category-theory}.

\bibitemdeclare{inproceedings}{pyzx}
\bibitem{pyzx}
\bibinfo{author}{Aleks \surnamestart Kissinger\surnameend} \&
  \bibinfo{author}{John \surnamestart van~de Wetering\surnameend}
  (\bibinfo{year}{2020}): \emph{\bibinfo{title}{{PyZX: Large Scale Automated
  Diagrammatic Reasoning}}}.
\newblock In \bibinfo{editor}{Bob \surnamestart Coecke\surnameend} \&
  \bibinfo{editor}{Matthew \surnamestart Leifer\surnameend}, editors: {\slshape
  \bibinfo{booktitle}{{\rm Proceedings 16th International Conference on}
  Quantum Physics and Logic, {\rm Chapman University, Orange, CA, USA., 10-14
  June 2019}}}, {\slshape \bibinfo{series}{Electronic Proceedings in
  Theoretical Computer Science}} \bibinfo{volume}{318},
  \bibinfo{publisher}{Open Publishing Association}, pp.
  \bibinfo{pages}{229--241}, \doi{10.4204/EPTCS.318.14}.

\bibitemdeclare{article}{kissinger2022simulating}
\bibitem{kissinger2022simulating}
\bibinfo{author}{Aleks \surnamestart Kissinger\surnameend} \&
  \bibinfo{author}{John \surnamestart van~de Wetering\surnameend}
  (\bibinfo{year}{2022}): \emph{\bibinfo{title}{Simulating quantum circuits
  with ZX-calculus reduced stabiliser decompositions}}.
\newblock {\slshape \bibinfo{journal}{Quantum Science and Technology}}
  \bibinfo{volume}{7}(\bibinfo{number}{4}), p. \bibinfo{pages}{044001},
  \doi{10.1088/2058-9565/ac5d20}.

\bibitemdeclare{misc}{chyp}
\bibitem{chyp}
\bibinfo{author}{Alex \surnamestart Kissinger\surnameend}
  (\bibinfo{year}{2023}): \emph{\bibinfo{title}{{G}it{H}ub - akissinger/chyp:
  {A}n interactive theorem prover for string diagrams --- github.com}}.
\newblock \bibinfo{howpublished}{\url{https://github.com/akissinger/chyp}}.

\bibitemdeclare{inproceedings}{lafont2024diagram}
\bibitem{lafont2024diagram}
\bibinfo{author}{Ambroise \surnamestart Lafont\surnameend}
  (\bibinfo{year}{2024}): \emph{\bibinfo{title}{A diagram editor to mechanise
  categorical proofs}}.
\newblock In: {\slshape \bibinfo{booktitle}{35es Journ{\'e}es Francophones des
  Langages Applicatifs (JFLA 2024)}}.

\bibitemdeclare{misc}{2023vyzx}
\bibitem{2023vyzx}
\bibinfo{author}{Adrian \surnamestart Lehmann\surnameend}, \bibinfo{author}{Ben
  \surnamestart Caldwell\surnameend}, \bibinfo{author}{Bhakti \surnamestart
  Shah\surnameend} \& \bibinfo{author}{Robert \surnamestart Rand\surnameend}
  (\bibinfo{year}{2023}): \emph{\bibinfo{title}{VyZX: Formal Verification of a
  Graphical Quantum Language}}.
\newblock \eprint{2311.11571}.

\bibitemdeclare{article}{magnus79}
\bibitem{magnus79}
\bibinfo{author}{Jan~R. \surnamestart Magnus\surnameend} \&
  \bibinfo{author}{H.~\surnamestart Neudecker\surnameend}
  (\bibinfo{year}{1979}): \emph{\bibinfo{title}{The Commutation Matrix: Some
  Properties and Applications}}.
\newblock {\slshape \bibinfo{journal}{The Annals of Statistics}}
  \bibinfo{volume}{7}(\bibinfo{number}{2}), pp. \bibinfo{pages}{381--394}.
\newblock \urlprefix\url{http://www.jstor.org/stable/2958818}.

\bibitemdeclare{inproceedings}{megacz2007coinductive}
\bibitem{megacz2007coinductive}
\bibinfo{author}{Adam \surnamestart Megacz\surnameend} (\bibinfo{year}{2007}):
  \emph{\bibinfo{title}{A coinductive monad for prop-bounded recursion}}.
\newblock In: {\slshape \bibinfo{booktitle}{Proceedings of the 2007 workshop on
  Programming languages meets program verification}}, pp.
  \bibinfo{pages}{11--20}, \doi{10.1145/1292597.1292601}.

\bibitemdeclare{inproceedings}{paykin2017qwire}
\bibitem{paykin2017qwire}
\bibinfo{author}{Jennifer \surnamestart Paykin\surnameend},
  \bibinfo{author}{Robert \surnamestart Rand\surnameend} \&
  \bibinfo{author}{Steve \surnamestart Zdancewic\surnameend}
  (\bibinfo{year}{2017}): \emph{\bibinfo{title}{{{QWIRE}}: A Core Language for
  Quantum Circuits}}.
\newblock In: {\slshape \bibinfo{booktitle}{Proceedings of the 44th ACM SIGPLAN
  Symposium on Principles of Programming Languages}}, \bibinfo{series}{{{POPL}}
  '17}, \bibinfo{publisher}{{Association for Computing Machinery}},
  \bibinfo{address}{{New York, NY, USA}}, pp. \bibinfo{pages}{846--858},
  \doi{10.1145/3009837.3009894}.
\newblock \urlprefix\url{https://jpaykin.github.io/papers/prz_qwire_2017.pdf}.

\bibitemdeclare{misc}{rand2018phantom}
\bibitem{rand2018phantom}
\bibinfo{author}{Robert \surnamestart Rand\surnameend},
  \bibinfo{author}{Jennifer \surnamestart Paykin\surnameend} \&
  \bibinfo{author}{Steve \surnamestart Zdancewic\surnameend}
  (\bibinfo{year}{2018}): \emph{\bibinfo{title}{Phantom {{Types}} for {{Quantum
  Programs}}}}.
\newblock
  \urlprefix\url{https://popl18.sigplan.org/event/coqpl-2018-phantom-types-for-quantum-programs}.
\newblock \bibinfo{note}{Talk at The Fourth International Workshop on Coq for
  Programming Languages (CoqPL '18)}.

\bibitemdeclare{incollection}{Selinger2010}
\bibitem{Selinger2010}
\bibinfo{author}{P.~\surnamestart Selinger\surnameend} (\bibinfo{year}{2010}):
  \emph{\bibinfo{title}{A Survey of Graphical Languages for Monoidal
  Categories}}.
\newblock In: {\slshape \bibinfo{booktitle}{New Structures for Physics}},
  \bibinfo{publisher}{Springer Berlin Heidelberg}, pp.
  \bibinfo{pages}{289--355}, \doi{10.1007/978-3-642-12821-9\_4}.

\bibitemdeclare{inproceedings}{coq-typeclasses}
\bibitem{coq-typeclasses}
\bibinfo{author}{Matthieu \surnamestart Sozeau\surnameend} \&
  \bibinfo{author}{Nicolas \surnamestart Oury\surnameend}
  (\bibinfo{year}{2008}): \emph{\bibinfo{title}{First-Class Type Classes}}.
\newblock In: {\slshape \bibinfo{booktitle}{Proceedings of the 21st
  International Conference on Theorem Proving in Higher Order Logics}},
  \bibinfo{series}{TPHOLs '08}, \bibinfo{publisher}{Springer-Verlag},
  \bibinfo{address}{Berlin, Heidelberg}, p. \bibinfo{pages}{278–293},
  \doi{10.1007/978-3-540-71067-7\_23}.

\bibitemdeclare{book}{hottbook}
\bibitem{hottbook}
\bibinfo{author}{The \surnamestart {Univalent Foundations Program}\surnameend}
  (\bibinfo{year}{2013}): \emph{\bibinfo{title}{Homotopy Type Theory: Univalent
  Foundations of Mathematics}}.
\newblock \bibinfo{publisher}{\url{https://homotopytypetheory.org/book}},
  \bibinfo{address}{Institute for Advanced Study}.

\bibitemdeclare{misc}{UniMath}
\bibitem{UniMath}
\bibinfo{author}{Vladimir \surnamestart Voevodsky\surnameend},
  \bibinfo{author}{Benedikt \surnamestart Ahrens\surnameend},
  \bibinfo{author}{Daniel \surnamestart Grayson\surnameend} et~al.:
  \emph{\bibinfo{title}{UniMath --- a computer-checked library of univalent
  mathematics}}.
\newblock \bibinfo{howpublished}{available at \url{http://unimath.org}},
  \doi{10.5281/zenodo.10849216}.
\newblock \urlprefix\url{https://github.com/UniMath/UniMath}.

\bibitemdeclare{misc}{vandewetering2020zxcalculus}
\bibitem{vandewetering2020zxcalculus}
\bibinfo{author}{John \surnamestart van~de Wetering\surnameend}
  (\bibinfo{year}{2020}): \emph{\bibinfo{title}{ZX-calculus for the working
  quantum computer scientist}}, \doi{10.48550/arXiv.2012.13966}.
\newblock \eprint{2012.13966}.

\end{thebibliography}

\end{document}